\documentclass[twoside,twocolumn]{article}

\usepackage[colorlinks]{hyperref}
\usepackage{breakurl}
\usepackage{adjustbox}
\usepackage{amsmath}

\usepackage{blindtext} 

\usepackage[sc]{mathpazo} 
\usepackage[T1]{fontenc} 
\usepackage{microtype} 

\usepackage[english]{babel} 

\usepackage[hmarginratio=1:1,top=32mm,columnsep=20pt]{geometry} 
\usepackage[hang, small,labelfont=bf,up,textfont=it,up]{caption} 
\usepackage{booktabs} 

\usepackage{lettrine} 

\usepackage{enumitem} 
\setlist[itemize]{noitemsep} 

\usepackage{abstract} 

\usepackage{titlesec} 
\renewcommand\thesection{\Roman{section}} 
\renewcommand\thesubsection{\roman{subsection}} 
\titleformat{\section}[block]{\large\scshape\centering}{\thesection.}{1em}{} 
\titleformat{\subsection}[block]{\large}{\thesubsection.}{1em}{} 

\usepackage{titling} 
\usepackage{cite}

\pretitle{\begin{center}\Huge\bfseries} 
\posttitle{\end{center}} 
\title{Indirect search for color octet electron at PWFA-LC} 
\author{%
\textsc{A. N. Akay}\\[1ex] 
\normalsize TOBB University of Economics and Technology - Ankara, TURKEY \\ 
\normalsize \href{mailto:aakay@etu.edu.tr}{aakay@etu.edu.tr} 
\and 
\textsc{U. Kaya}\thanks{Corresponding author} \\[1ex] 
\normalsize  Ankara University, Department of Physics - Ankara, TURKEY \\ 
\normalsize \href{mailto:umit.kaya@cern.ch}{umit.kaya@cern.ch} 
\and
\textsc{S. Turkoz}\\[1ex] 
\normalsize Ankara University, Department of Physics - Ankara, TURKEY \\ 
\normalsize \href{mailto:turkoz@science.ankara.edu.tr}{turkoz@science.ankara.edu.tr} 
}
\date{\today} 
\renewcommand{\maketitlehookd}{%
\begin{abstract}
Indirect search for color octet electron at Plasma Wake Field Accelerator- Linear Collider (PWFA-LC) has been considered. It is shown that color octet electron masses can be probed up to 17.7 TeV at PWFA-LC with three years integrated luminosity. In addition, the compositeness scale can be probed up to 38.5 TeV.
\end{abstract}
}

\begin{document}

\maketitle


\section{Introduction}

The predictions of the Standard Model (SM) has been verified by numerous experiments. In 2012, ATLAS and CMS experiments found a new particle with a mass of about 125 GeV
\cite{ATLASHiggs,CMSHiggs,CMSHiggs2}. More precise measurements \cite{ATLASprec,ATLASprec2,CMSprec,CMSprec2} have established that all observed properties of the new particle are consistent with the SM Higgs boson. However, there are a lot of problems, which have not solutions in the SM framework. In order to solve these problems, many beyond the SM models have been proposed. Among them composite models are favoured by historical arguments: periodic table of chemical elements was clarified by Rutherford experiment, hadron inflation has resulted in quark model. According to the compositeness, SM quarks and leptons should be made of more fundamental constituents. These constituents are called as preon by Pati and Salam.\\
Leptoquarks, excited leptons, excited quarks, dileptons, diquarks and leptogluons (color octet leptons) are predicted by composite models. Leptoquarks, excited quarks and excited leptons are included in the research programs of ATLAS and CMS experiments, however the color octet electron is not directly investigated in these experiments. Color octet electron is strongly interacting partners of SM leptons.\\
Up to this time, many experimental researches on $e_{8}$ have been done. First experimental bound on color octet electron ($e_{8}$), $M_{e_{8}} > 86$ GeV, presented in \cite{PDG} is based on  CDF search \cite{CDF}. Leptogluons with mass up to 200 GeV was exluded by  D0 experiment \cite{Hewett-Rizzo}. H1 search for $e_{8}$ excluded the compositeness scale $\Lambda < 3$ TeV for $M_{e_{8}} > 100$ GeV and $\Lambda < 240$ GeV for $M_{e_{8}} > 250$ GeV \cite{H1, H1-2}. Although the LEP experiments did not perform a direct search for leptogluons, low limits for excited lepton masses, namely 103.2 GeV \cite{PDG}, certainly is valid for $l_{8}$, too. Finally, reinterpratation of CMS results on leptoquark searches performed in \cite{Netto} leads to the strongest current limit on the $e_{8}$ mass, $M_{e_{8}} > 1.2 - 1.3$ TeV. \\
There are a number of phenomenological studies on $l_{8}$ production at TeV colliders. For example, production of leptogluons at the LHC has been analyzed in  \cite{Celikel, Mandal, Netto,Zuridov, Mandal2}. Resonant production of leptogluons at ep and $\mu$p colliders was considered in  \cite{Celikel2, Sahin, Sahin2} and \cite{Cheung} respectively. Indirect production of leptogluons at ILC and CLIC has been studied in  \cite{Akay}. On the other hand, considering IceCube PeV events  \cite{Aartsen}, color octet neutrinos may be the source of these extraordinary events  \cite{Akay2}.\\
In this paper, we consider indirect production of color octet electron at PWFA-LC. Main parameters of PWFA-LC are discussed in Section II. In Section III, we present the interaction lagrangian and indirect production cross-section of color octet electron. Signal and background analysis have been considered in Section IV. Finally, we summarize our results and conclude in Section V.

\section{Main Parameters of PWFA-LC}

Beam driven plasma wake field technology made a great progress
for linear accelerators recently. This method enables an electron beam
to obtain high gradients of energy even only propagating through small
distances compared to the radio frequency resonance based accelerators \cite{PWFA}. In other words, more compact linear accelerators can be
built utilizing PWFA to obtain a specified beam energy. In Table \ref{tab.PWFA}, main collider parameters of PWFA-LC  are listed. 

\begin{table}[!htbp]
\caption{Main parameters of PWFA-LC.}
\label{tab.PWFA}
\begin{center}
\scalebox{0.85}{%
\begin{tabular}{|l|l|l|}
\hline
Beam Energy (GeV) & 5000 \\
 \hline
Peak Luminosity ($10^{34} cm^{-2} s^{-1}$) & 6.27 \\
\hline
Particle per bunch ($10^{10}$) & 1.00 \\
\hline
Norm. Horiz. emittance ($\mu$m) &  0.01\\
\hline
Norm. Vert. emittance (nm) & 0.350\\
\hline
Horiz. $\beta^{*}$ amplitude function at IP (mm) &11.0 \\
\hline
Vert. $\beta^{*}$ amplitude function at IP (mm) & 0.099\\
\hline
Horiz. IP beam size (nm) & 106\\
\hline
Vert. IP beam size (nm) & 59.8\\
\hline
Bunches per beam & 1\\
\hline
Repetition rate (Hz) & 5000\\
\hline
Beam power at IP (MW) & 40.0\\
\hline
Bunch spacing ($10^{4}$ns) & 20.0\\
\hline
Bunch length (mm) & 0.02\\
\hline
\end{tabular} }
\end{center}
\end{table}
\section{Interaction Lagrangian and Production Cross-Section}
The interaction Lagrangian of leptogluons with the corresponding lepton and gluon is given by  \cite{lag1,lag2,lag3}:

\begin{equation}
\label{eq.1}
L=\frac{1}{2\Lambda}\Sigma_{l}\lbrace{\bar{l_{8}}}g_{s}G^{\alpha}_{\mu\nu}\sigma^{\mu\nu}(\eta_{L}l_{L}+\eta_{R}l_{R})+h.c.\rbrace
\end{equation}

where $G^{\alpha}_{\mu\nu}$
is field strength tensor for gluon, index $\alpha$=1, 2 , ..., 8 denotes the color, $g_{s}$ is gauge coupling, $\eta_{L}$ and $\eta_{R}$ are the chirality factors, $l_{L}$ and $l_{R}$ denote left and right spinor components of lepton, $\sigma^{\mu\nu}$ is the antisymmetric tensor and $\Lambda$ is the compositeness scale, which is taken to be equal to $M_{e_{8}}$. The leptonic chiral invariance implies $\eta_{L}$ $\eta_{R}$ =0. For numerical calculations we use the CalcHep program \cite{Calchep}.

\begin{figure}[!htbp]
\centering
\includegraphics[width=0.35\textwidth]{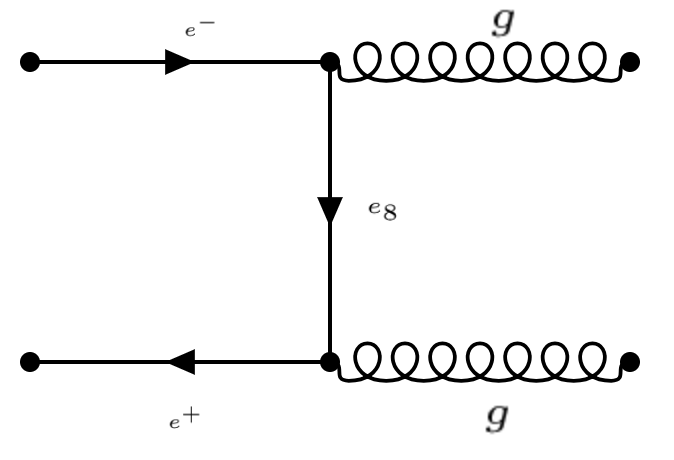}
\caption{Feynman diagram of indirect production of $e_{8}$.}
\label{feyndiag}
\end{figure}

\begin{figure}[!htbp]
\centering
\includegraphics[width=0.45\textwidth]{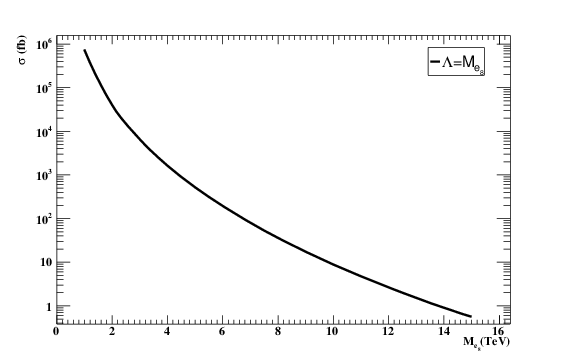}
\caption{Indirect production crossection of $e_{8}$ at PWFA-LC.}
\label{xsecPWFA}
\end{figure}

Signal process is $e^{-} e^{+} \rightarrow g g$ and corresponding Feynman diagram is shown in Figure \ref{feyndiag}. In Figure  \ref{xsecPWFA}, we present indirect production cross-section of $e_{8}$ at PWFA-LC.

\section{Signal and Background Analysis}

\begin{figure}[!htbp]
\centering
\includegraphics[width=0.45\textwidth]{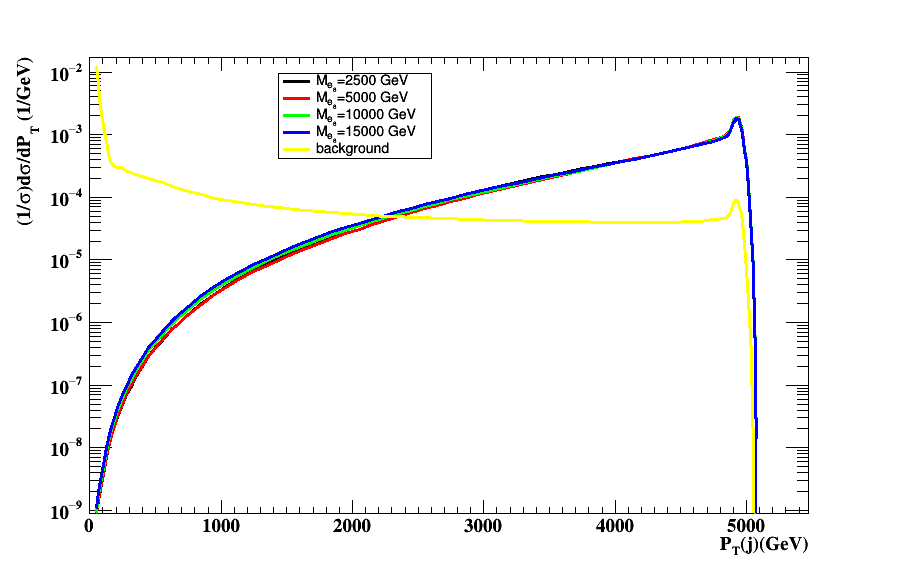}
\caption{$P_{T}$ distribution of final state jets at PWFA-LC.}
\label{PWFAPt}
\end{figure}

\begin{figure}[!htbp]
\centering
\includegraphics[width=0.45\textwidth]{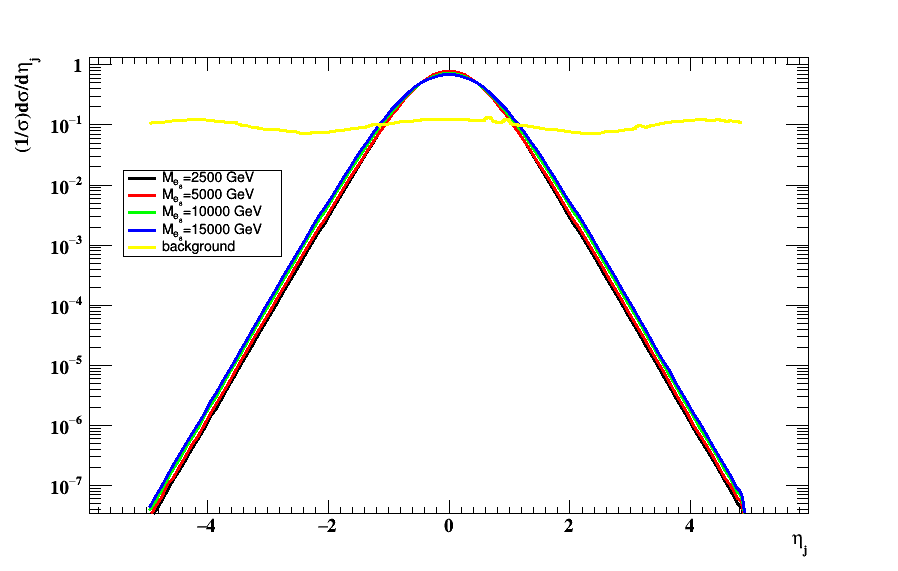}
\caption{$\eta$ distribution of final state jets at PWFA-LC.}
\label{PWFAeta}
\end{figure}

\begin{figure}[!htbp]
\centering
\includegraphics[width=0.45\textwidth]{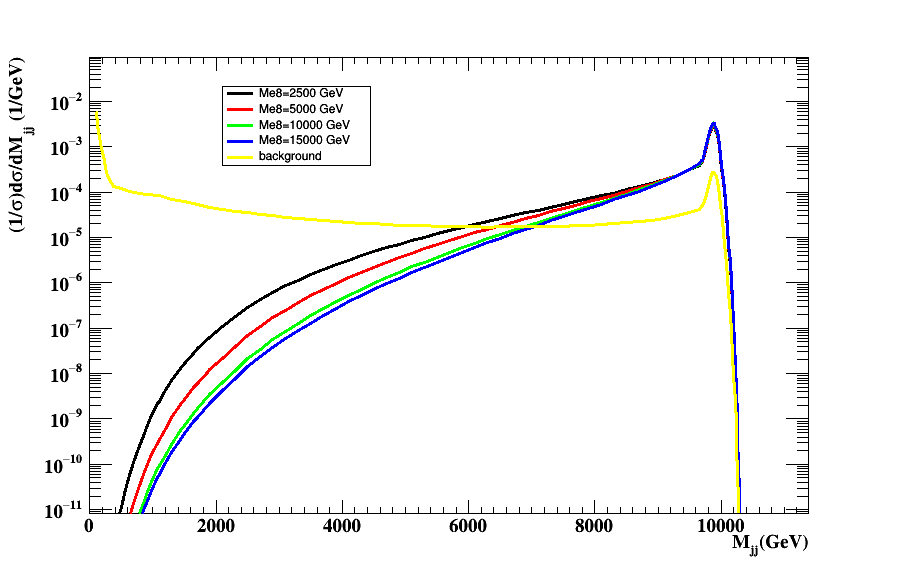}
\caption{$M_{jj}$ distribution of final state jets at PWFA-LC.}
\label{PWFAinv}
\end{figure}

As we mentioned previously our signal process is $e^{-} e^{+} \rightarrow g g$ and corresponding background process is $e^{-} e^{+}\rightarrow \gamma,Z\rightarrow j j$ ($j= u,\bar{u}, d, \bar{d}, c, \bar{c}, s, \bar{s}, b, \bar{b}$). In order to differentiate signal and background, we compare transverse momentum ($P_{T}$), pseudo-rapidity ($\eta$) and invariant mass distribution of final state jets of signal and background processes. In Figure \ref{PWFAPt}, \ref{PWFAeta} and \ref{PWFAinv} we show $P_{T}$, $\eta$ and $M_{jj}$ distributions of final state jets at PWFA-LC.

\begin{figure}[!htbp]
\centering
\includegraphics[width=0.45\textwidth]{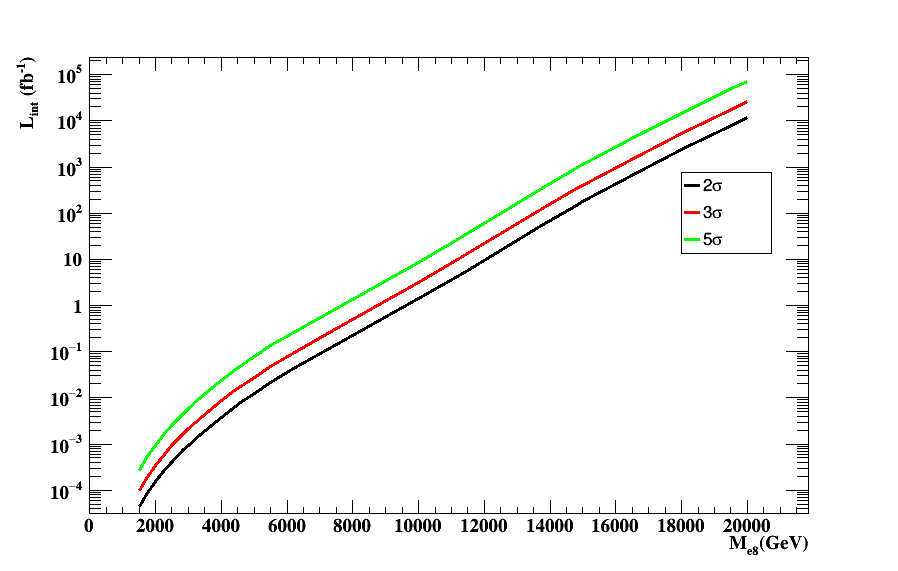}
\caption{The necessary integrated luminosity for the indirect observation of $e_{8}$ at PWFA-LC.}
\label{lumineeds}
\end{figure}

In order to reduce background we determine $P_{T}$ and $\eta$ cut and mass window values from the kinematical distributions. We  apply $P_{T} > 2350 GeV$, $\mid\eta\mid < 1.0$ cuts and $M_{jj} > 7000 GeV$ for  PWFA-LC. We use the the formula given in Eq.(\ref{eq.SS}) for statistical significance:

\begin{equation}
\label{eq.SS}
SS= \dfrac{\sigma_{s}}{\sqrt{\sigma_{s}+\sigma_{b}}}\sqrt{L_{int}}
\end{equation}

where $\sigma_{s}$ is signal cross-sections, $\sigma_{b}$ denotes background cross-sections and $L_{int}$ is integrated luminosity. In Table \ref{tab.masslimits}, reachable $e_{8}$ mass value for $2\sigma$ (exlusion), $3\sigma$ (observation) and $5\sigma$ (discovery) limits at PWFA-LC are given.  In Figure \ref{lumineeds}, necessary luminosities as a function of $e_{8}$ mass for $2\sigma$, $3\sigma$ and $5\sigma$ are given. \\
So far, we assumed the compositeness scale equal to the mass of color octet electron: $\Lambda=M_{e_{8}}$. In fact, this scale may be different from the mass of the color octet electron. For this reason, we estimate the limits of compositeness scale when the color octet electron mass is 5000 GeV. We show reachable compositeness scale ($\Lambda$) values for $2\sigma$, $3\sigma$ and $5\sigma$ limits at PWFA-LC in Table \ref{tab.lambdalimits}.

\begin{table}[!htbp]
\caption{ Reachable $e_{8}$ mass values (in TeV) at PWFA-LC .}
\label{tab.masslimits}
\begin{center}
\scalebox{0.85}{%
\begin{tabular}{|l|l|l|l|l|}
\hline
Colliders & years & 5$\sigma$ & 3$\sigma$ & 2$\sigma$ \\
\hline
PWFA-LC &  1 & 14.4 & 15.5 & 16.4  \\
\hline
PWFA-LC & 3 & 15.6 & 16.7 & 17.7 \\
\hline
\end{tabular} }
\end{center}
\end{table}

\begin{table}[!htbp]
\caption{Reachable $\Lambda$ values (in TeV) at PWFA-LC ($M_{e_{8}}$= 5 TeV).}
\label{tab.lambdalimits}
\begin{center}
\scalebox{0.85}{%
\begin{tabular}{|l|l|l|l|l|}
\hline
Colliders & years & 5$\sigma$ & 3$\sigma$ & 2$\sigma$ \\
\hline
PWFA-LC &  1 & 26.4 & 30.2 & 33.4  \\
\hline
PWFA-LC & 3 & 30.5 & 34.7 & 38.5  \\
\hline
\end{tabular}}
\end{center}
\end{table}

\section{Conclusion}

In this paper, we have studied indirect production of color octet electron at PWFA-LC. We determined 2$\sigma$ exclusion, 3$\sigma$ observation and 5$\sigma$ discover limits of e8 at PWFA-LC. PWFA-LC will give opportunity to exclude, observe and discovery of color octet electron up to 16.4 TeV, 15.5 TeV and 14.4 TeV respectively with one year collider operation. These numbers become 17.7 TeV, 16.7 TeV and 15.6 TeV respectively with three year collider operation. If $e_{8}$ is discovered with 5 TeV mass, then PWFA-LC with three year integrated luminosity will give opportunity to probe compositeness scale up to 38.5 TeV.

\section*{Acknowledgement}
Authors are grateful to Saleh Sultansoy for useful discussions.


\begin{thebibliography}{99}

\bibitem[1]{ATLASHiggs}
ATLAS Collaboration (2012),
\newblock {\em Phys. Lett. B}, 716, 1.

\bibitem[2]{CMSHiggs}
CMS Collaboration (2012),
\newblock {\em Phys. Lett. B}, 716, 30.

\bibitem[3]{CMSHiggs2}
CMS Collaboration (2013),
\newblock {\em J. High Energy Phys.}, 06, 081.
  
\bibitem[4]{ATLASprec}
ATLAS Collaboration (2013),
 \newblock {\em Phys. Lett. B}, 726, 88.
  
\bibitem[5]{ATLASprec2}
ATLAS Collaboration (2013),
\newblock {\em Phys. Lett. B}, 726, 120.
  
\bibitem[6]{CMSprec}
CMS Collaboration (2015),
\newblock {\em Eur. Phys. J. C}, 75, 212.
  
\bibitem[7]{CMSprec2}
CMS Collaboration (2015),
\newblock {\em Phys. Rev. D}, 92, 012004. 
  
\bibitem[8]{PDG}
Olive K. A., et al. (Particle Data Group) (2104),
\newblock {\em Chin. Phys. C}, 38, 090001. 
  
\bibitem[9]{CDF}
Abe F., et al. (CDF Collaboration) (1989),
\newblock {\em Phys. Rev. Lett.}, 63, 1447. 
  
\bibitem[10]{Hewett-Rizzo}
Hewett J. L. and Rizzo T. G. (1997),
\newblock {\em Phys. Rev. D}, 56, 5709. 
  
\bibitem[11]{H1}
Abt I., et al. (H1 Collaboration) (1993),
\newblock {\em Nucl. Phys. B}, 396, 3. 
  
\bibitem[12]{H1-2}
Ahmed T., et al. (H1 Collaboration) (1994),
\newblock {\em Z. Phys. C}, 64, 545. 

\bibitem[13]{Celikel}
Celikel A., Kantar M. and Sultansoy S. (1998),
\newblock {\em Phys. Lett. B}, 443, 359.

\bibitem[14]{Mandal}
Mandal T. and Mitra S. (2013),
\newblock {\em Phys. Rev. D}, 87, 095008.
  
\bibitem[15]{Netto}
Goncalves-Netto D., et al. (2013),
\newblock {\em Phys. Rev. D}, 87, 094023.   

\bibitem[16]{Zuridov}
Jelinski T. and Zhuridov D. (2015),
\newblock {\em Acta Phys. Pol. B}, 46, 2185.
  
\bibitem[17]{Mandal2}
Mandal T., Mitra S. and Seth S. (2016),
\newblock {\em Phys. Lett. B},758, 219-25.
   
\bibitem[18]{Celikel2}
Celikel A. and Kantar M. (1998),
\newblock {\em Turk. J. Phys.}, 22, 401.
  
\bibitem[19]{Sahin}
Sahin M., Sultansoy S. and Turkoz S.(2010),
\newblock {\em Phys. Lett. B}, 689, 172.
  
\bibitem[20]{Sahin2}
Sahin M. (2014),
\newblock {\em Acta Phys. Pol. B}, 45, 1811.
  
\bibitem[21]{Cheung}
Cheung K. (2000),
\newblock {\em AIP Conf. Proc.}, 542, 160.
  
\bibitem[22]{Akay}
Akay A. N., et al. (2011),
\newblock {\em EPL}, 95, 31001.
  
\bibitem[23]{Aartsen}
Aartsen M. G., et al. (2014),
\newblock {\em Phys. Rev. Lett.}, 113, 101101.
  
\bibitem[24]{Akay2}
Akay A. N., et al. (2015),
\newblock {\em Int. J. Mod. Phys. A}, 30, 1550163.                      

\bibitem[25]{PWFA}
J-P. Delahaye, et al. (2014),
\newblock {\em Proceedings of the Fifth International Particle Accelerator
Conference, Dresden, Germany}, p. 3791.

\bibitem[26]{lag1}
Celikel A. and Kantar M. (1998),
\newblock {\em Turk. J. Phys.}, 22, 10.

\bibitem[27]{lag2}
Sahin M., Sultansoy S. and Turkoz S. (2010),
\newblock {\em Phys. Lett. B}, 689, 172.

\bibitem[28]{lag3}
Particle Data Group (Nakamura K. et al. (2010),
\newblock {\em J. Phys. G}, 37, 075021.

\bibitem[29]{Calchep}
Pukhov A.,Belyaev A. and Christensen N. (2013),
\newblock {\em Computer Physics Communications}, 184, 1729-1769.

\end{thebibliography}
\end{document}